# Using data mining to explore calmodulin bibliography

Jacques Haiech* and Marie-Claude Kilhoffer

CNRS UMR7242 BSC, ESBS, 300 Bd Sébastien Brant, CS 10413, 67412 ILLKIRCH cedex

* Corresponding author: haiech@unistra.fr

**i. Running Head**: using data mining in the calcium field

**ii. Summary**

In this chapter, we present a strategy and the technics to approach a scientific field from a set of articles gathered from the bibliographic database, Web of Science. The strategy is based on methods developed to analyze social networks. We illustrate the use of such strategy in studying the calmodulin field. Such method allows to structure a huge number of articles when writing a review, to detect the key opinion leaders in a given field and to locate his own research topic in the landscape of the themes deciphered by our own community.

We show that the free software VosViewer may be used without knowledge in computing science and with a short learning period.

## 1. Introduction

The number of publications has increased exponentially in the last 40 years. It is becoming almost impossible to perform a bibliographic analysis using a traditional approach by gathering all new articles and analyzing them. We need to amplify our intelligences to be able to tackle the tsunami of new publications and to get a coherent landscape describing the evolution of a given scientific field.

Such an approach has been developed in scientometry, a discipline that aims to analyze science by means of quantitative and qualitative indicators. This discipline is linked to the humanities and therefore, the methods and strategies developed by scientometrists have not yet perfused in Biology. The strategy has been used to analyze the emergence of synthetic biology but by a team of sociologists[1].

In this article, we aim to present some of the strategies and tools used in scientometry and as an example, to show how such an approach may help to understand the function of calmodulin.

The number of articles containing calmodulin in the title is higher than 10000 and over 40000 when seeking for calmodulin in the abstract or the keywords. As a human being, it is almost impossible to scan this full set of articles and to extract a coherent and logical set of knowledge.

Data mining is a set of technics allowing to analyze data from different points of views in order to extract useful knowledge, mainly by analyzing the relations between the different data or by detecting common patterns.

Such analysis may be performed on a huge set of data and data mining is therefore often associated with Big Data.

In our case, the number of papers published on calmodulin cannot be considered as enormous, but it is big enough to escape our human abilities. We will use the study of calmodulin to illustrate the capabilities of data mining and the strategies used to extract knowledge from an unstructured set of publications. Such strategies and technics may be used in any scientific field.

We have realized that the use of this strategy by students allows them to better approach a new field of knowledge. Moreover, they can take into account the temporal evolution of the field and to tackle it via the most significant publications.

## 2. Materials and definitions

### 2.1. Corpus:

A corpus is a set of articles and is defined by three parameters: the name of a data base, a request and a date. In our case, we will use either PUBMED [2] or Web of Science[3] as bibliographic data base. Both data bases allow to download the result of a request under several possible formats **(Figure 1**).

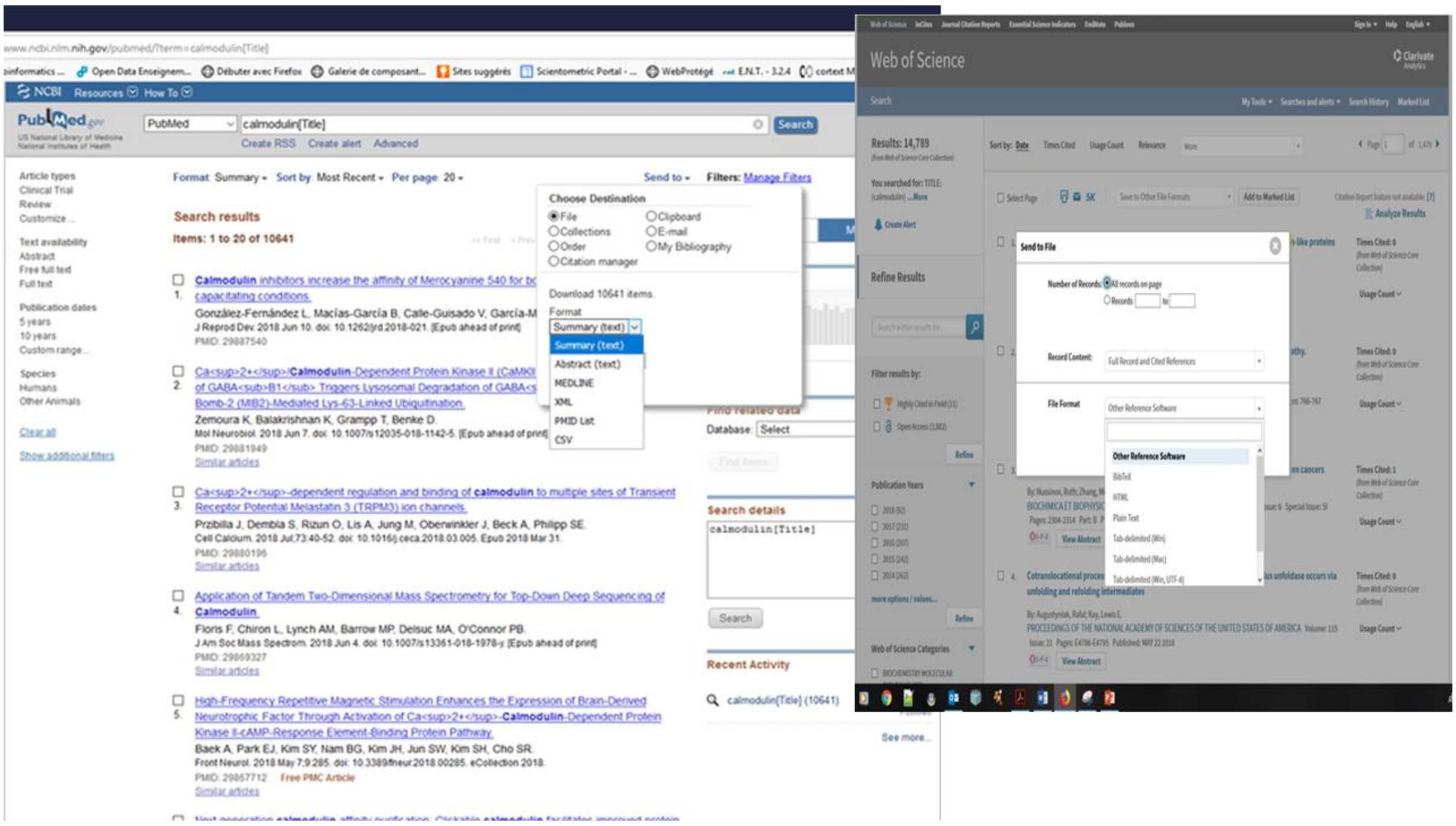

**Figure 1:** To create the corpus, we use either the PUBMED database (left panel) or the Web of Science database (right panel) using the simple request “calmodulin” in the title field. The download of the corpus is done in PUBMED under the format MEDLINE and in WoS under the format text.

### 2.2. Analyzing and visualizing:

In this paper, we will use Vosviewer version 1.6.8 as the tool to analyze and visualize our data [4].

The use of PUBMED files allows to create maps of researchers based on coauthorship links and maps of terms based on co-occurrences in the same publications of keywords either provided by the authors or by the curators of the bibliographic databases.

On top of the previous maps, the use of Web of Science files allows one to easily create maps of scientific publications, scientific journals, researchers, or research organizations based on either co-citation relations (i.e., multiple items being cited by the same publication) or bibliographic coupling relations (i.e., multiple items citing the same publication).

VosViewer contains an algorithm able to extract terms from titles and abstracts using natural language processing tools. Basically, a term is a suite of adjective(s) and noun(s) finishing by a noun. The algorithm is working only on English text. A map of term is built by linking two terms when they are present in the same publication. As we will see in methods and results, we will see that each available tool allows to explore an aspect of the gathered corpus.

## 3. Methods and Results

### 3.1. The general strategy in data mining is composed of three steps (Figure 2):

1. Building the corpus by extracting the data from any source, by loading and by transforming the data including the possibility to cure the data

1) Analyzing the corpus by using tools allowing either to detect common patterns in the corpus or between subsets of the corpus or to detect relationships between specific properties of items of the corpus. In a bibliographic corpus, an item is an article and properties of an article are authors, title, abstract, year of publication, ….
2) Visualizing the results from the previous analyzes using either tools to represent clusters or graphs.

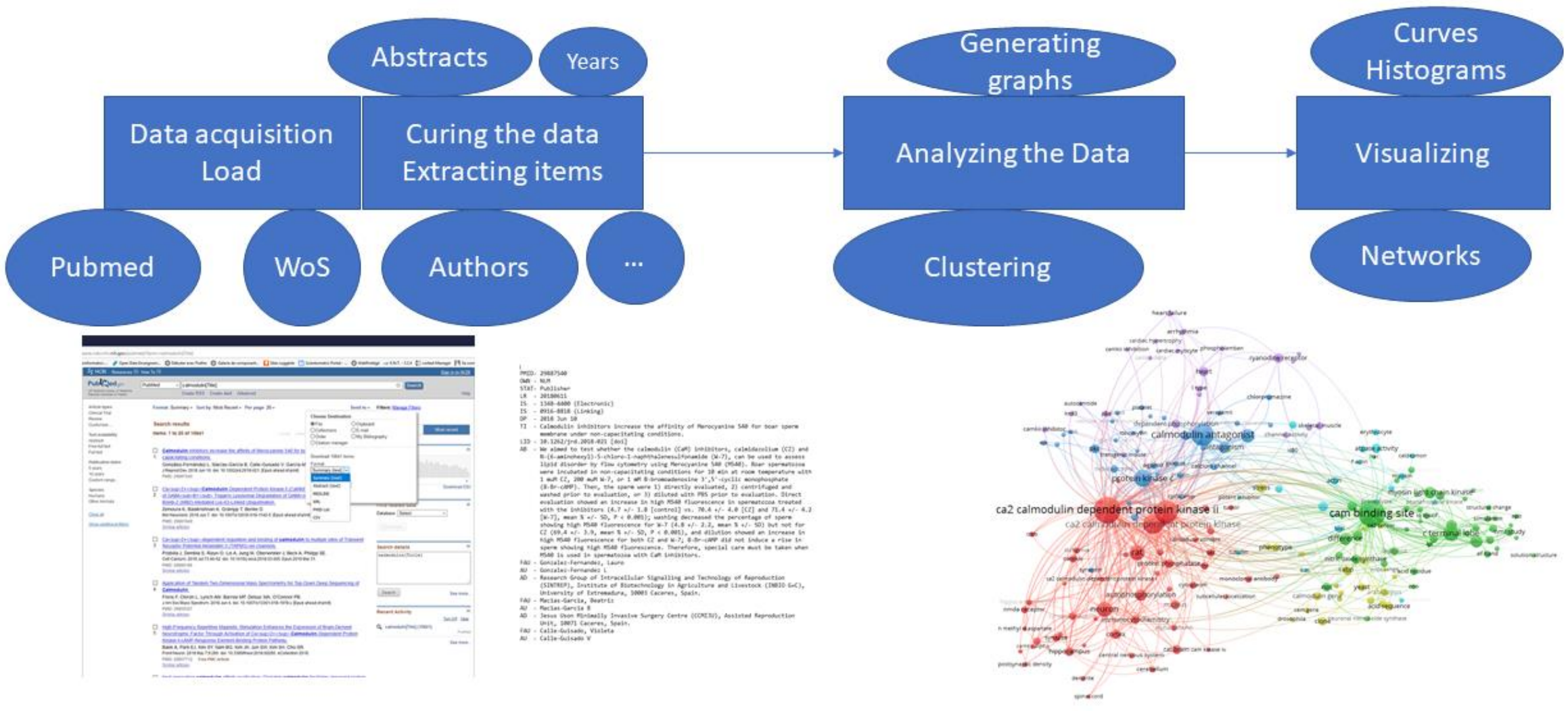

**Figure 2:** This scheme illustrates the strategy of data mining when using bibliographic data.

To illustrate our method, we will tackle the study of calmodulin, the stereotypical calcium binding protein involved in the modulation of numerous cellular events[5, 6].

To build the corpus, we will use at first a very simple request, namely looking for calmodulin in the title of the publications. We obtain in PUBMED 10641 publications and in Web of Science 14789 publications.

The difference between the two corpus is in part due to the presence of meeting abstracts and notes in the Web of Science database. We obtain 10389 items when we select only articles and reviews in the Web of Science corpus.

In the following part of the paper, we will use the complete corpus obtained from the Web of Science database (14789 items) onJune12$^{th}$ ,2018.

## 3.2. What are the main themes and their temporal evolutions in the calmodulin field?

To answer this question, after loading the corpus in the Vosviewer software, we use a tool allowing to create a map based on the text from titles and abstracts of the publications. Vosviewer may tag adjectives and nouns, extract terms composed of adjectives and nouns finishing by a noun and compute a score of relevance for each term. However, to improve the significance of the term set, it is needed to iteratively build a thesaurus file to remove non-significant terms in our studied field and to pinpoint the putative synonyms (see the thesaurus file as supplemental material S1). Such file is a text file with two columns separated by a tabulation. The two columns are with the heading "label" and "replace by".

Two terms are linked if they appear in the same publication. The strength of the link is the number of publications where the two terms co-occur. It is a parameter that we set to 50. Then, we choose to select the 200 most relevant terms. Vosviewer may draw the network named the term map. The default parameters given by Vosviewer were used to draw this map (**Figure 3**).

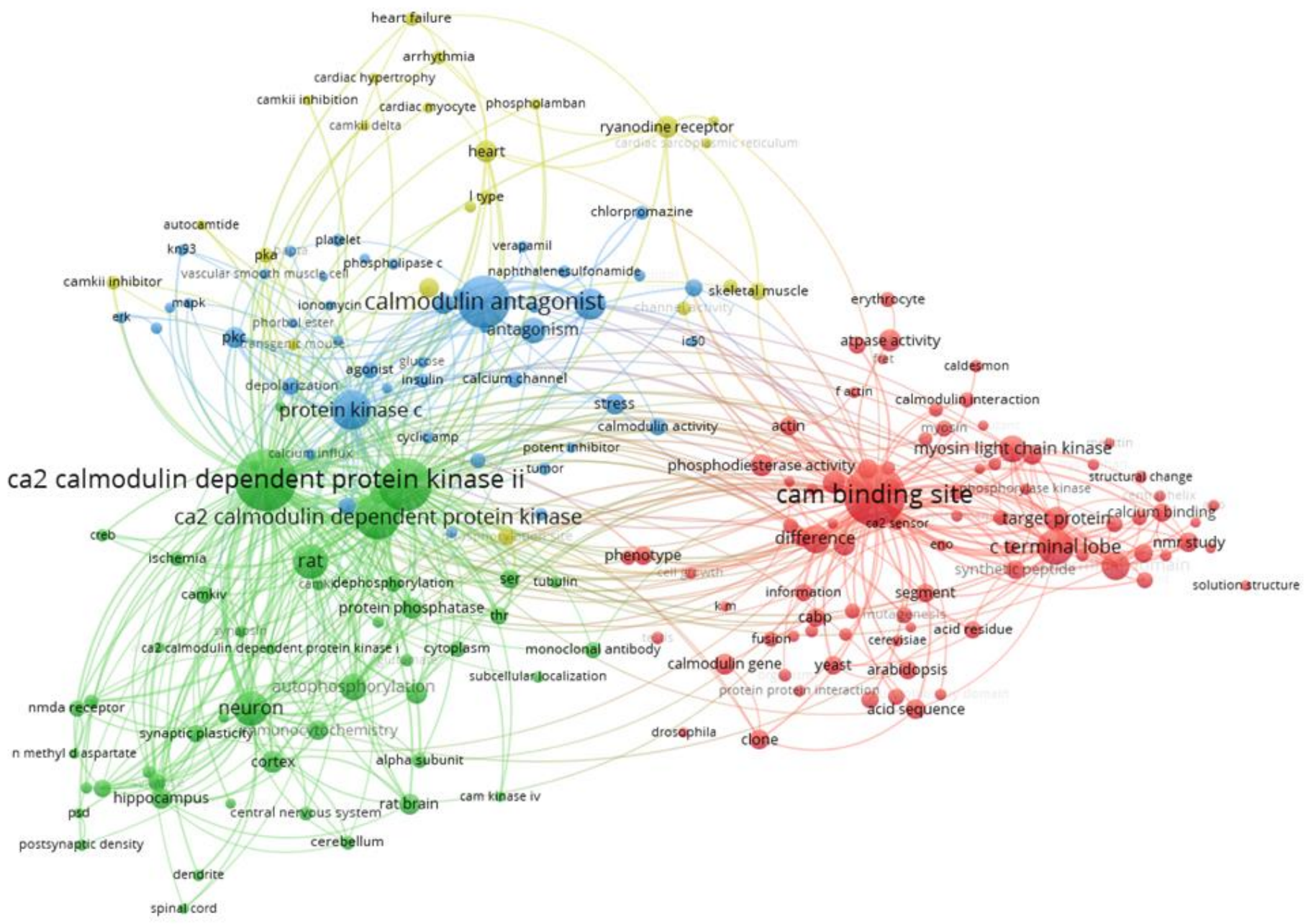

**Figure 3:** Term map created with Vosviewer version 1.6.8 by extracting significant terms from abstract and titles. The links are drawn when two terms appear together in more than 50 items.The 200 most significant terms were kept using a thesaurus file to remove general terms in the field or by regrouping synonyms (Thesaurus file as supplemental materials)

To perform the clustering, VosViewer uses a smart local moving[7] algorithm that we may optimize using two parameters (resolution and min.cluster size under the analysis signet). Using resolution equal to 1 and min.cluster size to 10, we obtain 4 clusters (**Figure 3**), one nucleated around calmodulin antagonist (blue color), one around calmodulin dependent protein kinase II (green color), one around calmodulin binding site (red color) and one around heart and modulation of calcium channel (yellow color). The cluster one is the most ancient and the cluster 4 the most recent. By increasing the resolution parameter, the initial clusters are subdivided into smaller ones. For instance, the cluster “calmoduline binding site” may be split into a cluster dealing with the characterization of calcium sites and another focusing on calmodulin binding peptides/targets.

At this point, we have one alternative:

1) Modifying the corpus to focus on one specific cluster or sub-cluster by adding terms to the initial request
2) Continuing with the initial corpus to explore other questions.

We will go along with the second part of the alternative.

## 3.3. How to find the main authors in the different topics and the key articles to read?

There are three maps that allow to answer those two questions of the section [8]:

1) The citation map creates a link between two publications when one of them cites the other,
2) The co-citation map links two publications when they are cited by the same article of the corpus,
3) The bibliographic coupling map creates a link between two articles of the corpus when both cites the same document[9].

Each of those maps may be build with publications or authors as nodes (Figure 4).

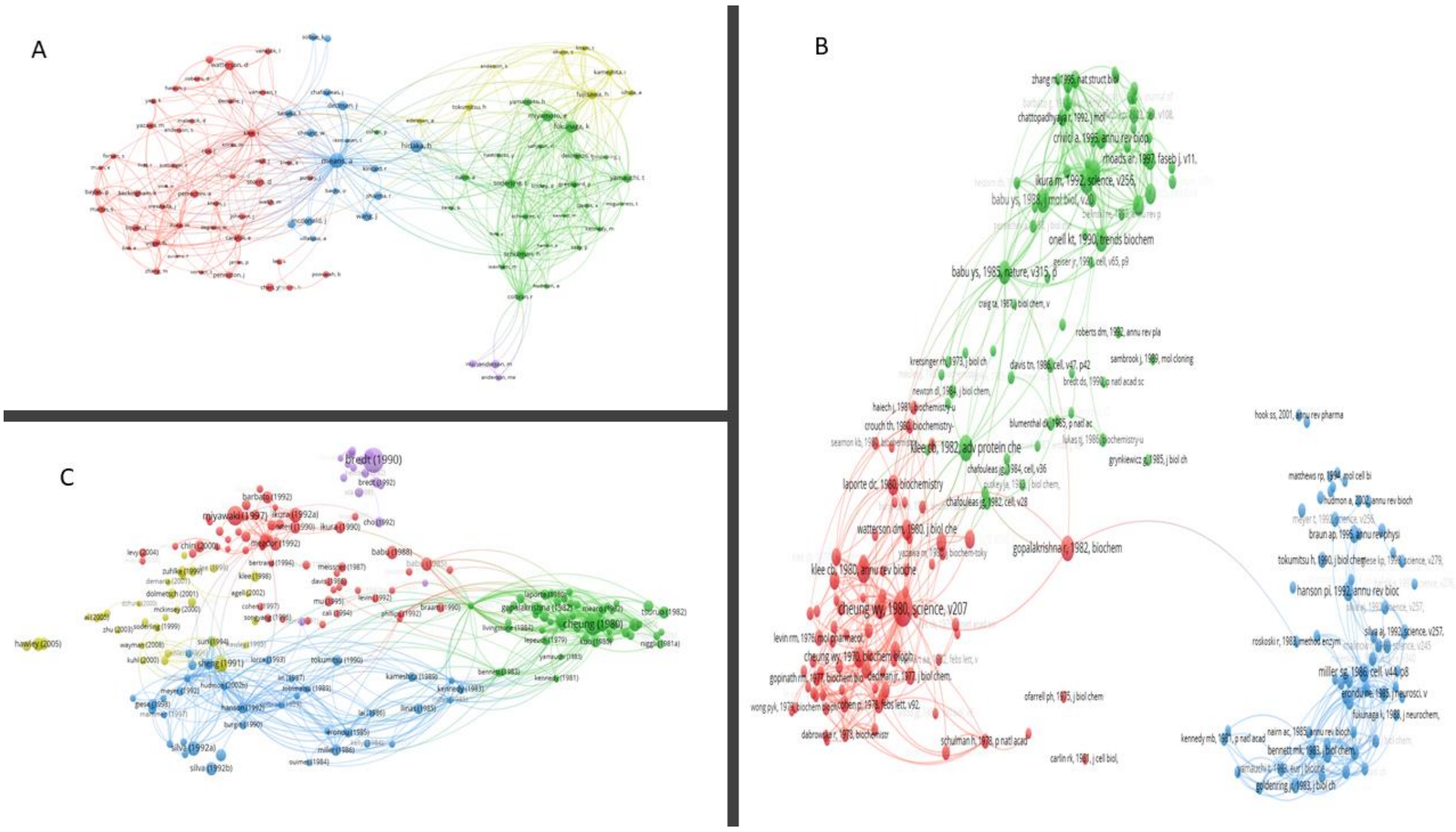

Figure 4: This figure presents the citation map (panel A), the co-citation map (panel B) and the bibliographic coupling map (panel C) obtained with Vosviewer version 1.6.8 and the parameters discussed in the section Methods and Results.

The citation map using authors as nodes (**Figure 4A**) has been built keeping the most significant 100 authors. This map pinpoints the five main scientific communities working in the calmodulin field. The first one (in red) does not show a central author nucleating the community although CB Klee is playing a central role, the second one (in green) appears to be organized around fives authors (E Miyamoto, K. Fukunaga, T Yamauchi, TR Soderling and H. Schulman) the third one (in blue) is clearly built around A Means, the fourth one appears to be centralized around H Fujisawa and I. Kameshita and finally, the fifth one is a small cluster of two authors, Anderson and Wu.

This first map allows to discriminate five scientific communities also pinpointing five main topics and their evolutions:

1) Cluster 1: This cluster gathers scientists that worked on calmodulin per se, namely the structure and the calcium binding properties of calmodulin.
2) Cluster 2: Scientists in this cluster mainly worked on the role of calmodulin kinases and specifically calmodulin kinase II in brain. The small cluster 5 is a spinoff of this cluster 2. The authors are working on calmodulin kinase II but in heart
3) Cluster 3: This cluster is structured around the work of AR Means on calmodulin kinase kinases and, recently, mainly on CaMKK2
4) Cluster 4: The work of the scientists in this cluster mainly deals with the couple calmodulin kinases and the phosphatases that reverse the phosphorylation of the calmodulin kinases.

The co-citation map has been constructed by keeping the 100 most significant articles (**Figure 4B**). In this map, we distinguished three clusters.

1) The first cluster presents papers published between 1970 up to 1982 and dealing with the primary characterization of calmodulin (purification, calcium binding properties, primary sequence, cloning the gene of calmodulin).
2) The second cluster gathers the publications describing mainly the 3D structure of calmodulin and calmodulin complexed with target proteins or peptides, for the period ranging from 1985 to 2003.
3) Finally, the third cluster deals with the calmodulin kinases.

Using this tool, it is possible to select for each cluster two publications that are the most cited. We may consider that the selected six papers constitute the fundamental publications to read (Table I). We may see that those six papers may be considered as an excellent starting point for a newcomer to enter the calmodulin field. Let notice that this choice of publications has been realized without any “à priori” or implication of knowledge from experts.

| Cluster | Publication | Reference |
|---|---|---|
| 1 | Calmodulin plays a pivotal role in cellular regulation.<br>Cheung WY. | [10] |
| 1 | Calmodulin.<br>Klee CB, Crouch TH, Richman PG. | [11] |
| 2 | Solution structure of a calmodulin-target peptide complex by multidimensional NMR.<br>Ikura M, Clore GM, Gronenborn AM, Zhu G, Klee CB, Bax A. | [12] |
| 2 | Target enzyme recognition by calmodulin: 2.4 A structure of a calmodulin-peptide complex.<br>Meador WE, Means AR, Quiocho FA. | [13] |
| 3 | Neuronal Ca2+/calmodulin-dependent protein kinases.<br>Hanson PI, Schulman H. | [14] |
| 3 | Regulation of brain type II Ca2+/calmodulin-dependent protein kinase by autophosphorylation: a Ca2+-triggered molecular switch.<br>Miller SG, Kennedy MB. | [15] |

Table 1: For each cluster from figure 4B, the two most cited reference were selected.

The third map is the bibliographic coupling map (Figure 4C). From a technical point of view, it is the exact opposite of the co-citation map. Co-citation map is a way to represent the “current paradigm” of the domain. It is a retrospective analysis. On the contrary, the bibliographic coupling map seems to be a prospective analysis and to detect the trends of the scientific domain. To build this map, we have selected the 200 most cited publications. Keeping the default parameters of Vosviewer for the clustering, we obtained 5 clusters:

1. The first cluster (in red) integrate work on the structure of calmodulin-complexes with target proteins or peptides and work on new calcium dyes using the properties of such complexes. The papers in this cluster were published around the year 2000.
2. The second cluster (in green) gathers works dealing with the biochemical purification and characterization of calmodulin and calmodulin interacting proteins. Small molecules interacting with calmodulin are used to perturb the system to decipher the molecular mechanism. The work in this cluster peaks around 1980.
3. The third cluster (in blue) is focusing on the CaM kinase II and specifically its role in the brain. Most of the work is done around 1990.
4. The fourth cluster (in yellow) is mainly dealing with CaM kinases in the cardiac tissue and/or the modulation of ion channels. The publications in this cluster are the most recent ones, peaking around 2005.
5. Finally, the fifth cluster (in violet) links most of the work on calmodulin dependent nitric oxide synthase mainly in endothelial cells, starting around 1990.

These clusters appear to describe the trends in the calmodulin field.

Combining our different maps, we can summarize the state of the art in the calmodulin field, the six publications to read for the beginners in the field and the putative trends (**Figure 5**).

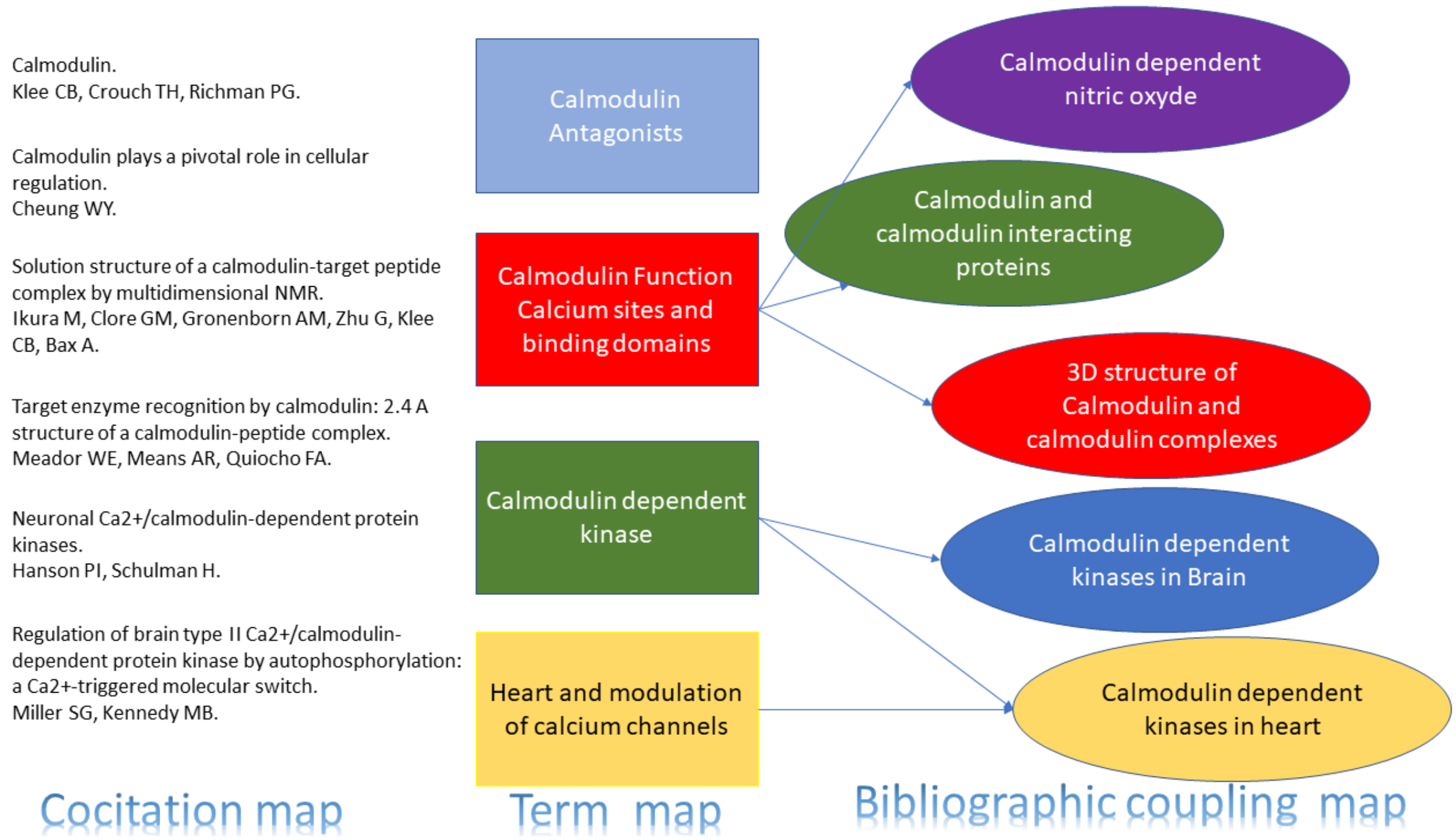

**Figure 5:** Combining the results from the different maps. Colors for the term map and the bibliographic coupling map are the same as those used in the figures 3 and 4 for the visualization of the different clusters.

Calmodulin has been discovered in 1968 as a calcium dependent regulator of phosphodiesterase and the name calmodulin was coined to the protein in 1977. At first, the scientific community aimed at purifying and characterizing the protein and cloning its genes. In parallel, calmodulin interacting proteins started to be identified and the molecular mechanism of interaction was investigated. Simultaneously, the interaction of calmodulin with small molecules was studied, mainly in Japan and triggered a subdomain of investigation on calmodulin antagonists.

The number of calmodulin interacting proteins rapidly increased up to several hundred. The 3D structure of calmodulin was solved first by X ray crystallography and then by NMR.These results were rapidly followed by the structure of calmodulin complexed to target proteins domains or synthetic calmodulin interacting peptidess.

Among the large family of calmodulin interacting proteins, the kinase family attracted many scientists. Calmodulin dependent kinase II became one of the most studied calmodulin target and its role was mainly investigated in brain function.

More recently, the discovery of mutations in calmodulin linked to cardiac pathologies, prompted the field exploring the role of calmodulin dependent enzymes in the regulation of cardiac physiology.

Finally, as the oxidative stress appeared to play a specific action in oncogenesis, calmodulin dependent nitric oxide synthase became a hot topic in the field.

Nowadays, as shown in the **Figure 6**, the number of publications with the term "calmodulin" in the title is decreasing.

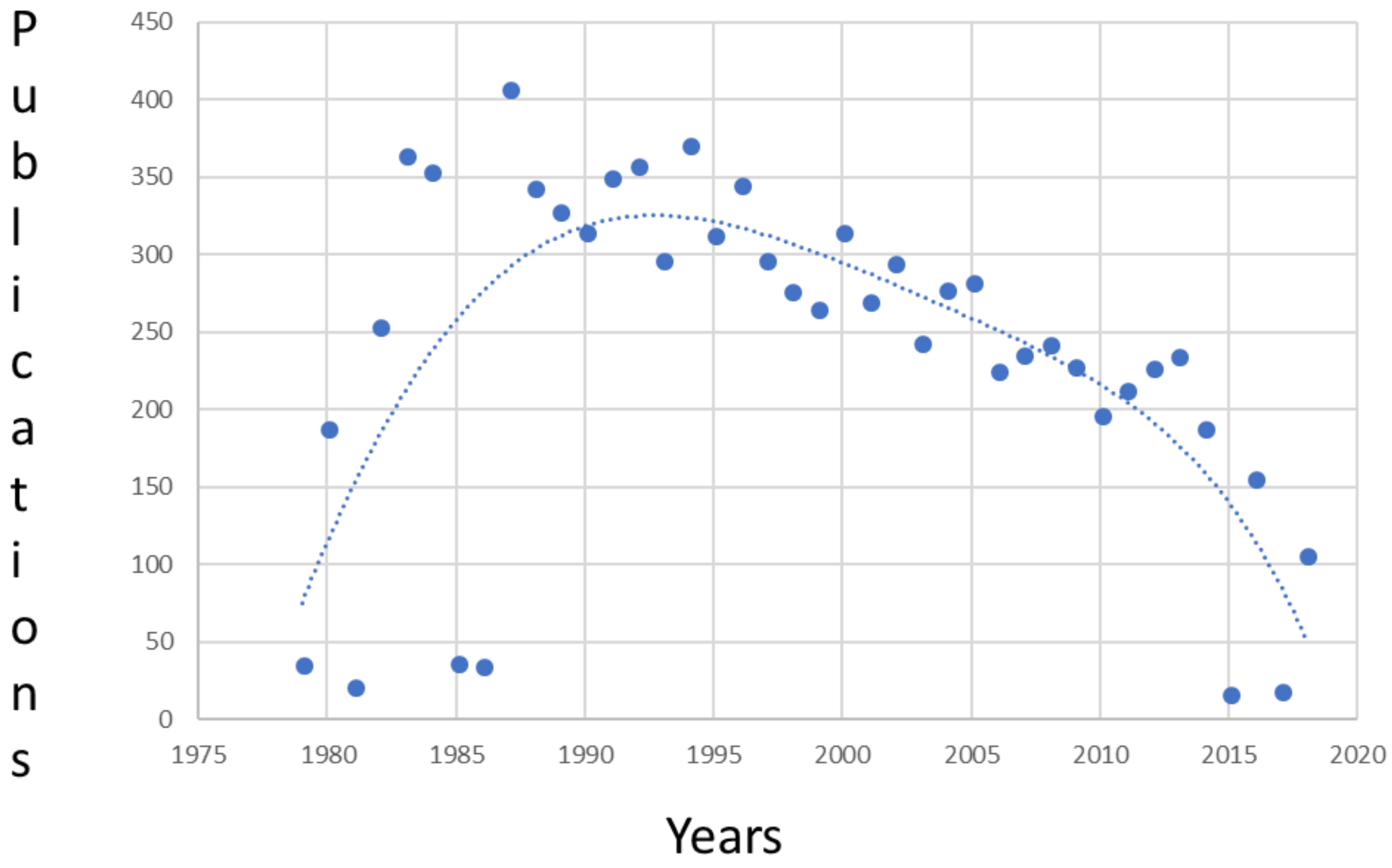

**Figure 6:** Number of publications per year with the term "calmodulin" in the title.

The number of publication was at the acme between 1985 and 1995. We may assume that nowadays, the scientific questions are more global and does not tackle the role of one given protein, as important as it can be.

In this chapter, we have shown how to use tools first developed in the field of social network analysis to tackle questions as diverse as:

1) What are the different subtopics of a given scientific field?
2) Who are the key opinion leaders in a given topic?
3) What are the main publications in a given subtopics?
4) How a scientific field has evolved upon time?

We have illustrated the use of such method in the frame of the evolution of the calmodulin field but this approach may be used in any field and may help to answer other questions such the ranking of different scientific institutes for a given scientific domains.

## 4. Notes

In this section, we will cite some limits of the strategy.

### 1. The Corpus:

It is important to work with large corpus (few hundred publications at least) in order to get meaningful and robust result.

On the other hand, it is important to polish the request to get a corpus specific of the studied domain. For instance, using a request seeking calmodulin in the abstract and not only in the title, will gather papers with all kind of calcium binding proteins and will not focus on calmodulin.

### 2. The thesaurus file:

The thesaurus file is important to get a significant network of terms when analyzing the different subtopics of a scientific field.

The aim is to remove all the terms that are often present in the abstract but does not bring any meaningful information. For instance, the terms of the request have to be removed or general term such as biology.

Building the thesaurus file is an iterative and subjective process. We have noticed that during this iterative process, removing one term may change the geometry of the network. However, the clustering of the terms are not profoundly modified.

### 3. The co-citation analysis

The co-citation analysis allows to retrospectively analyze the field. However, when the same type of technics are used in the field, we end up with technical papers that are not pertinent for a specific field.

For instance, the paper describing the technics to make a western blot or to measure the protein concentration must be removed when doing a co-citation analysis.

## References

1. Raimbault B, Cointet JP, Joly PB (2016) Mapping the Emergence of Synthetic Biology. PLoS One 11(9): p. e0161522
2. Vellay SG, Latimer NE, Paillard G (2009) Interactive text mining with Pipeline Pilot: a bibliographic web-based tool for PubMed. Infect Disord Drug Target 9(3): p. 366-374
3. Li, K., J. Rollins, and E. Yan, *Web of Science use in published research and review papers 1997-2017: a selective, dynamic, cross-domain, content-based analysis.* Scientometrics, 2018. **115**(1): p. 1-20.
4. van Eck, N.J. and L. Waltman, *Software survey: VOSviewer, a computer program for bibliometric mapping.* Scientometrics, 2010. **84**(2): p. 523-538.
5. Berchtold, M.W. and A. Villalobo, *The many faces of calmodulin in cell proliferation, programmed cell death, autophagy, and cancer.* Biochim Biophys Acta, 2014. **1843**(2): p. 398-435.
6. Haiech, J., et al., *Mutant analysis approaches to understanding calcium signal transduction through calmodulin and calmodulin regulated enzymes.* Adv Exp Med Biol, 1990. **269**: p. 43-56.
7. Waltman, L. and N.J. van Eck, *A smart local moving algorithm for large-scale modularity-based community detection.* The European Physical Journal B, 2013. **86**(11): p. 471.
8. Klavans, R. and K.W. Boyack, *Which Type of Citation Analysis Generates the Most Accurate Taxonomy of Scientific and Technical Knowledge?* Journal of the Association for Information Science & Technology, 2017. **68**(4): p. 984-998.
9. Weinberg, B.H., *Bibliographic coupling: A review.* Information Storage and Retrieval, 1974. **10**(5): p. 189-196.
10. Cheung, W.Y., *Calmodulin plays a pivotal role in cellular regulation.* Science, 1980. **207**(4426): p. 19-27.
11. Klee, C.B., T.H. Crouch, and P.G. Richman, *Calmodulin.* Annu Rev Biochem, 1980. **49**: p. 489-515.
12. Ikura, M., et al., *Solution structure of a calmodulin-target peptide complex by multidimensional NMR.* Science, 1992. **256**(5057): p. 632-8.
13. Meador, W.E., A.R. Means, and F.A. Quiocho, *Target enzyme recognition by calmodulin: 2.4 A structure of a calmodulin-peptide complex.* Science, 1992. **257**(5074): p. 1251-5.
14. Hanson, P.I. and H. Schulman, *Neuronal Ca2+/calmodulin-dependent protein kinases.* Annu Rev Biochem, 1992. **61**: p. 559-601.

15. Miller, S.G. and M.B. Kennedy, *Regulation of brain type II Ca2+/calmodulin-dependent protein kinase by autophosphorylation: a Ca2+-triggered molecular switch.* Cell, 1986. **44**(6): p. 861-70.